\documentclass[argument]{aastex62}
\usepackage{amsmath}
\usepackage{blindtext}
\graphicspath{{./}{figures/}}

\shorttitle{SWMF Simulations of Tidally Locked Exoplanets}
\shortauthors{Bagheri et al.}

\begin{document}

\title{Impacts of Tidal Locking on Magnetospheric Energy Input to Exoplanet Atmospheres}

\correspondingauthor{Fatemeh Bagheri}
\email{fatemeh.bagheri@nasa.gov}

\author{Fatemeh Bagheri}
\affil{NASA Goddard Space Flight Center,
Greenbelt, MD, USA}
\affil{The University of Texas at Arlington,
Arlington, TX, USA}

\author{Alex Glocer}
\affil{NASA Goddard Space Flight Center, 
Greenbelt, MD, USA}

\author{Ramon E. Lopez}
\affil{The University of Texas at Arlington,
Arlington, TX, USA}


\begin{abstract}

\noindent We investigate the effect of planetary corotation on energy dissipation within the magnetosphere-ionosphere system of exoplanets. Using MHD simulations, we find that tidally locked exoplanets have a higher cross-polar cap potential (CPCP) compared to fast-rotating planets with the same magnetic field strength, confirming previous studies. Our simulations show that for a given interplanetary magnetic field, an increase in corotation period leads to a higher CPCP. Notably, this difference in CPCP between tidally locked and rotating planets persists across a range of solar wind conditions, including extreme environments such as those experienced by hot Jupiters. Furthermore, we observe that variations in corotation have little impact on CPCP for Earth-sized planets. These results underscore the significance of both corotation dynamics and planetary size in understanding how exoplanets interact with their stellar environments.
\end{abstract}


\section{Introduction} \label{sec:intro}

\noindent Over the past few decades, the discovery of exoplanets has revolutionized our understanding of planetary systems and their diversity. Through a combination of observational techniques such as transit, radial velocity measurements, and direct imaging, astronomers have detected thousands of exoplanets spanning a wide range of sizes, compositions, and orbital characteristics \citep{wright2012frequency, knutson2014featureless, mandel2002analytic, seager2003unique, gaudi2012microlensing, bagheri2019detection, bagheri2024infrared, traub2010direct, guyon2005exoplanet, bagheri2024exploring}. These discoveries have challenged traditional theories of planet formation and evolution, leading to new insights into the prevalence of different types of planets, their potential habitability, and the conditions necessary for life to arise. \\ 

\noindent Early exoplanet catalogues were dominated by close-in exoplanets \citep{perryman2018exoplanet}. Close-in planets (semi-major axis a $< 0.05$ au) constitute a special subset of the exoplanetary population. Since it is unclear whether in-situ formation occurs, the current orbital and physical characteristics of these planets provide essential constraints on their past evolution and formation process\citep{jackson2008tidal}. The proximity of these exoplanets to their host stars leads to gravitational forces that gradually synchronize the planet's rotation with its orbit, resulting in a tidally locked state. Consequently, these planets are expected to permanently present one side to the star, with the opposite side in perpetual darkness, while a transitional zone, known as the terminator, experiences perpetual twilight.\\
\\
Potentially habitable terrestrial exoplanets may also be tidally locked. They are frequently observed orbiting M dwarf stars, such as Proxima Centauri b \citep{anglada2016terrestrial} and TRAPPIST-1e \citep{gillon2017seven}. These stars, characterized by their lower temperature and brightness relative to the Sun, require that planets orbit closer to them to be in the habitable zone. Hence, the planet is subjected to more potent tidal forces from the host star compared to Earth's interaction with the Sun, potentially leading to the synchronization of the planet's rotation rate with its orbital period \citep{pierrehumbert2019atmospheric}.\\
\\
While tidal locking should be a common occurrence in many exoplanetary systems, its influence on magnetospheric dynamics and subsequent interaction with stellar winds remains an area of limited exploration \citep{bagheri2024mhd}. Understanding the intricate interplay between tidal locking and the interaction of magnetospheres with stellar winds is crucial for several reasons. The magnetosphere acts as a protective barrier, shielding a planet's atmosphere from erosion caused by the intense radiation and charged particles emitted by its parent star's stellar wind. Consequently, any alterations or modulations in the structure of the magnetosphere resulting from tidal locking could significantly impact the habitability and evolutionary trajectory of exoplanets. Tidal locking also significantly affects the generation of auroral radiation due to asymmetrical conductance in the ionosphere \citep{zarka2001magnetically, seager2002constraining}. Thus, accurate estimation of exoplanetary radio emissions in MHD simulations requires consideration of the effects of tidal locking and exoplanet rotation. A slower rotation rate might weaken the dynamo effect, resulting in weaker magnetic fields and smaller magnetospheres. However, recent studies \citep{zuluaga2013influence} have indicated a non-trivial relationship between rotation period and magnetic properties, suggesting that tidally locked planets could still exhibit intense magnetic fields and extended magnetospheres, possibly with larger polar cap areas.\\
\\
\citet{bagheri2024mhd} used the GAMERA MHD code \citep{zhang2019gamera} to investigate energy dissipation in the magnetospheres of various exoplanet types. They found that tidally locked planets exhibit higher Cross Polar Cap Potential (CPCP) compared to fast-rotating planets. However, their study did not explain the underlying reasons for this difference in CPCP values. The present study builds on their work by exploring the impact of tidal locking on magnetospheric dynamics and its interaction with stellar winds, using the Space Weather Modeling Framework (SWMF).

\section{Magneto-Hydro-Dynamic Simulation}

\noindent A global magnetohydrodynamic (MHD) simulation, also known as a global MHD model, is an adequate theoretical/computational framework to study the behavior and dynamics of magnetized plasmas on large scales, encompassing entire astrophysical objects or systems. Global MHD simulations are often used to study complex astrophysical phenomena, such as the dynamics of stellar atmospheres, the interaction between planets and their magnetospheres with stellar winds, the formation and evolution of galaxies, and the behavior of accretion disks around black holes or young stellar objects. Several general-purpose MHD codes have been developed over the past decade, including but not limited to ZEUS \citep{stone1992zeus}, VAC \citep{toth1996general}, BATS-R-US code \citep{powell1999solution}, LFM \citep{lyon2004lyon}, PENCIL \citep{dobler2006magnetic}, RAMSES \citep{fromang2006high}, PLUTO \citep{mignone2007pluto}, FLASH code \citep{dubey2008challenges}, Athena \citep{stone2008athena}, Nirvana \citep{ziegler2008nirvana}, AstroBEAR \citep{cunningham2011astrobear} and GAMERA \citep{zhang2019gamera}. Of these MHD codes, BATS-R-US, LFM, and GAMERA have been applied to a wide range of magnetospheric research and are therefore well tested and appropriate to studying the magnetospheres of exoplanets.
\\
\\
In this paper, we use BATS-R-US (Block-Adaptive Tree Solar-wind Roe-type Upwind Scheme), for investigating the tidal locked effects. BATS-R-US is a widely used MHD model employed to study the Earth’s magnetosphere \citep{powell1999solution,toth2012adaptive}. BATS-R-US is often run as the magnetospheric component of the SWMF. The SWMF allows for different physical models to be coupled together to more accurately represent the space environment system \citep{toth2012adaptive}. For example, in magnetospheric simulations, the global magnetospheric component is often run with an ionospheric electrodynamics component \citep{ridley2004ionospheric}. In some cases an inner magnetospheric component is also included \citep{glocer2013crcm+} or a component capturing the escape of ionospheric plasma to the magnetosphere \citep{glocer2009modeling}. In this study we only include the global magnetosphere and ionosphere electrodynamics components. Among MHD codes, one of the pioneering features of the BATS-R-US model is its ability to dynamically and adaptively refine the computational cells, thus increasing resolution in areas of interest and decreasing it elsewhere. 
The BATS-R-US model has shown promising results in studying space weather of Earth and is widely used to examine the magnetospheres of exoplanets and their interactions with stellar winds. This model has provided valuable insights into how stellar winds affect exoplanetary atmospheres and magnetic environments. Numerous studies have applied this model to better understand the dynamics of exoplanet magnetospheres, including their role in protecting planets from atmospheric escape due to stellar wind exposure [e.g., \citet{cohen2009interactions, cohen2011dynamics, cohen2014magnetospheric, cohen2015interaction,  cohen2018exoplanet, garraffo2016space, dong2017habitable, airapetian2020impact,  chin2024role}].
\\
\\
In this study, we couple the Global Magnetosphere (GM) model based on BATS-R-US with the Ionospheric Electrodynamics (IE) model (Ridley Ionosphere Model). For the ionospheric parameters, we adopt a constant Pedersen conductivity model in all simulations. The use of a constant Pedersen conductivity model in our simulations is a reasonable approach, particularly for exploring broad-scale dynamics and first-order interactions. Jupiter-like planets, with their large magnetospheres and strong intrinsic magnetic fields, experience significant energy input from both the stellar wind and internal processes. In our simulations, the primary goal is to understand the overall coupling between the magnetosphere and the ionosphere, rather than to capture every local variation in ionospheric conditions. The constant Pedersen conductivity model simplifies the complex spatial and temporal variations in ionospheric conductivity, allowing for a clearer focus on the global transfer of energy from the solar wind into the magnetosphere. By assuming a uniform conductivity, this model helps to isolate the key drivers of magnetospheric dynamics, such as solar wind pressure and magnetic field interactions, which are crucial for understanding large-scale energy deposition and current flow in the system. While more detailed models of conductivity would provide finer insights into regional effects, the constant model provides a practical and computationally efficient tool for simulating the overall energy exchange in the magnetosphere-ionosphere system, especially when the primary aim is to compare the global magnetospheric behavior and large-scale energy transfer in tidally locked vs. rotating planets.
\\
\\
All simulations presented hereby involve a northward magnetic dipole for the planets and a southward IMF for the solar wind. This setup was chosen because energy transfer into the magnetosphere-ionosphere system is maximized when the IMF is antiparallel to the planetary dipole. This allows for a clearer comparison between tidally locked and rotating planets. When the Interplanetary Magnetic Field (IMF) is southward, magnetic reconnection primarily occurs at the dayside magnetopause, where the solar wind's magnetic field opposes the planet’s magnetic field. This opposition facilitates a highly efficient and rapid reconnection process, enabling the solar wind to transfer energy into the planet’s magnetosphere. The result is a surge of energy that can drive intense geomagnetic storms, destabilizing the magnetosphere and triggering the release of energy, which can lead to the formation of auroras. The reconnection at the magnetopause during a southward IMF is direct and swift, producing dramatic effects on the planet’s space weather. In contrast, when the IMF is northward, the magnetic reconnection primarily occurs in the lobes of the Earth's magnetotail, far from the Earth’s immediate vicinity. In this case, the IMF and the Earth's magnetic field are aligned, which inhibits efficient reconnection at the magnetopause. The reconnection in the lobes occurs more slowly and transfers energy into the magnetosphere in a less direct way. While this reconnection still allows some solar wind energy to penetrate the magnetosphere, it does so at a reduced rate, leading to weaker geomagnetic activity. The auroras associated with northward IMF tend to be less intense, and geomagnetic storms are much less likely, as the reconnection is slower and less energetic than during a southward IMF.

\section{Tidally locked vs. Fast-Rotating planets}\label{sec:0.4au}

\noindent Expanding on the  prior research of \citet{bagheri2024mhd}, we seek to expound upon the impact of rotational speed on energy dynamics within planetary magnetospheres. This investigation involves comparing scenarios with distinct corotational speeds: a rapidly rotating planet akin to Jupiter and a planet which is rotating synchronously with its orbital period (i.e., tidally locked). To address this, we will employ simulations analogous to those used in our previous study, focusing on a planet situated at approximately 0.4 au from a Sun-like star. For this distance, solar wind parameters are derived from actual measurements near Mercury \citep{diego2020properties}. These include a southward IMF with a strength of approximately $32$ nT, a solar wind temperature of about $0.2$  M K, a solar wind speed of $406$ km/s, and a solar wind density of $44 ~cm^{-3}$. These values provide a realistic basis for analyzing the interaction between the planet's magnetosphere and the solar wind under different rotational conditions. The solar wind parameters used in this section are summarized in Table \ref{tab:sw04au}. 
\begin{center}
Key Input Parameters for Simulations in Section \ref{sec:0.4au}
\\
\begin{tabular}{lllcc}
\hline\hline
corresponds to the orbital distance [au]	&	 0.4   \\
\hline
$\rho_{sw} ~[\text{cm}^{-3}]$	&	44	        \\
$v_{sw} ~[\text{km/s}]$	        &	406	        \\
$B_{\text{IMF}} ~[\text{nT}]$	&	32	        \\
tidal corotation period [hour]  &	2220        \\
\hline
\end{tabular} \label{tab:sw04au}
\end{center}

\begin{figure}[ht!]
\centering
\includegraphics[width=120mm]{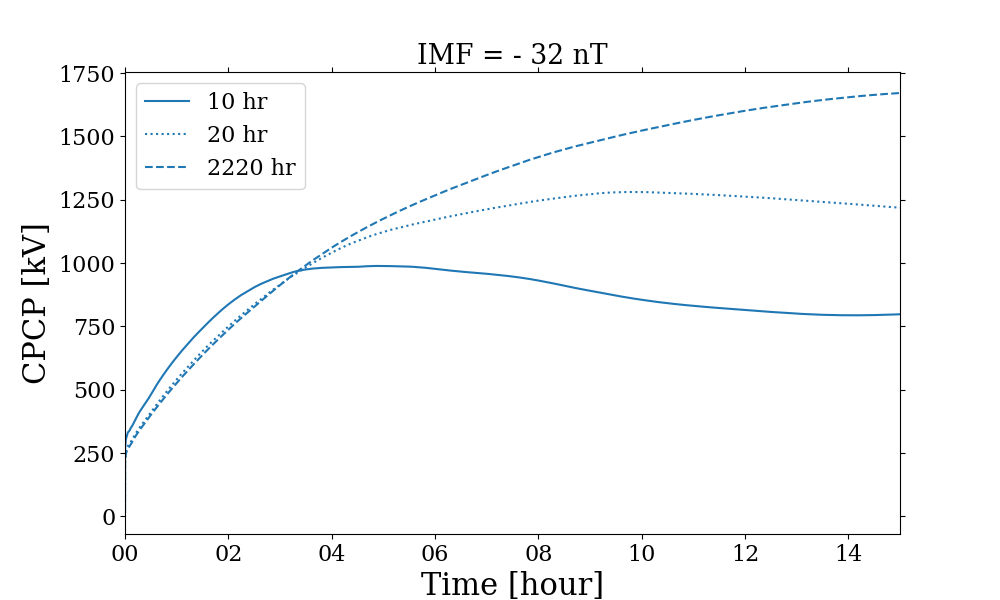}
\caption{Cross Polar Cap Potential for a tidally locked planet at 0.4 au (2220 hours corotation period), a slow-rotating (20 hours), and a fast-rotating planet (10 hours) in 15 hours of simulations.
\label{fig:tid_vs_rot_cpcp}}
\end{figure}

\noindent The magnetosphere's inner boundary is set to 5.5 planetary radii ($R_p$), and the upstream stellar wind boundary is located at 64 $R_p$. The downstream magnetotail outer boundary is positioned at 448 $R_p$. For simplicity, we assume that Jupiter's dipole magnetic moment aligns with its rotation axis and points southward, mirroring Earth's magnetic configuration. All planetary parameters, such as mass, radius, magnetic field strength, and angular velocity, are kept consistent with those of Jupiter. During the simulations, we maintain a constant ionospheric Pedersen conductance of approximately 105 using Equation 27 from \citep{nichols2016stellar} and a zero Hall conductance to standardize our analysis across different rotational speeds.
\\
\\
Similar to our study with the GAMERA MHD code \citep{bagheri2024mhd}, the SWMF simulations also indicate different behavior for the CPCP and different pattern for Field Aligned Currents (FACs) in the tidally locked planets. Illustrated in Figure \ref{fig:tid_vs_rot_cpcp}, the CPCP exhibits continuously increases, ultimately peaking at approximately twice that of the CPCP observed in the fast-rotating planet configuration with an equivalent magnetic field. 

\begin{figure}
\centering
\includegraphics[scale=0.2]{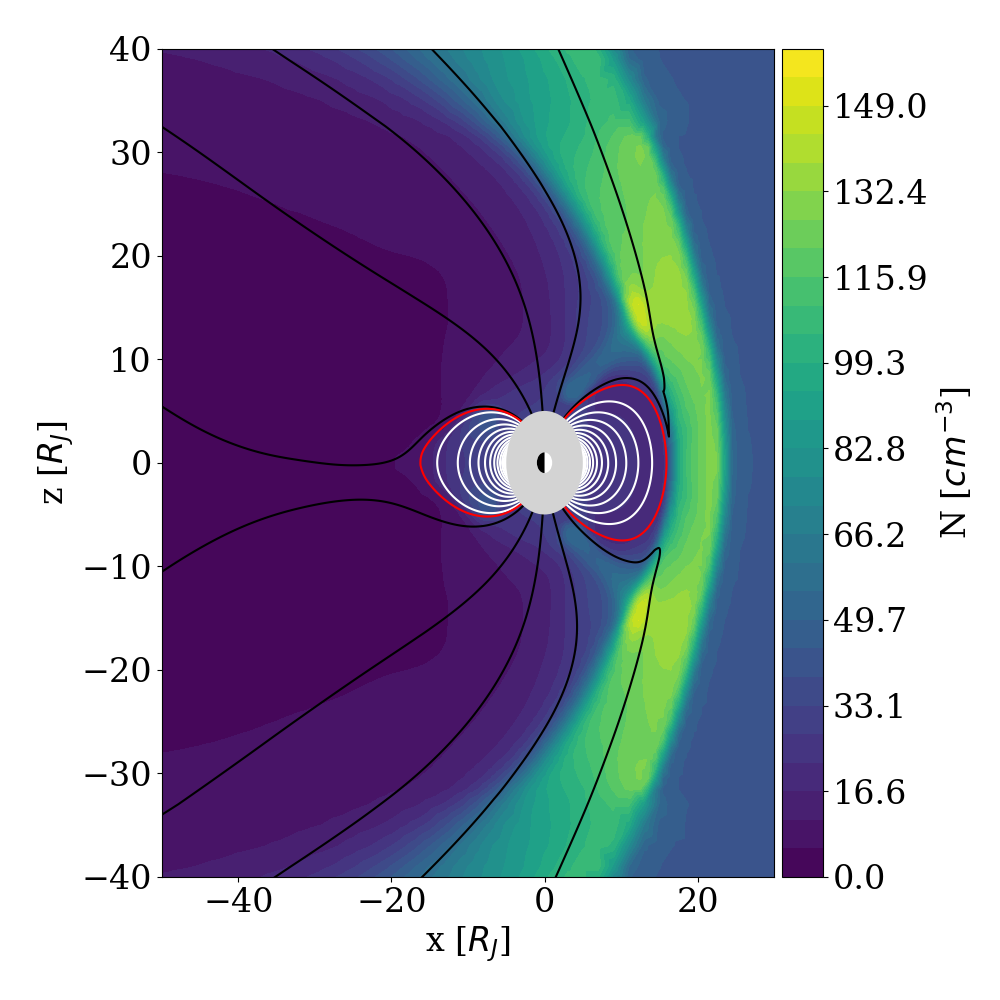}
\includegraphics[scale=0.2]{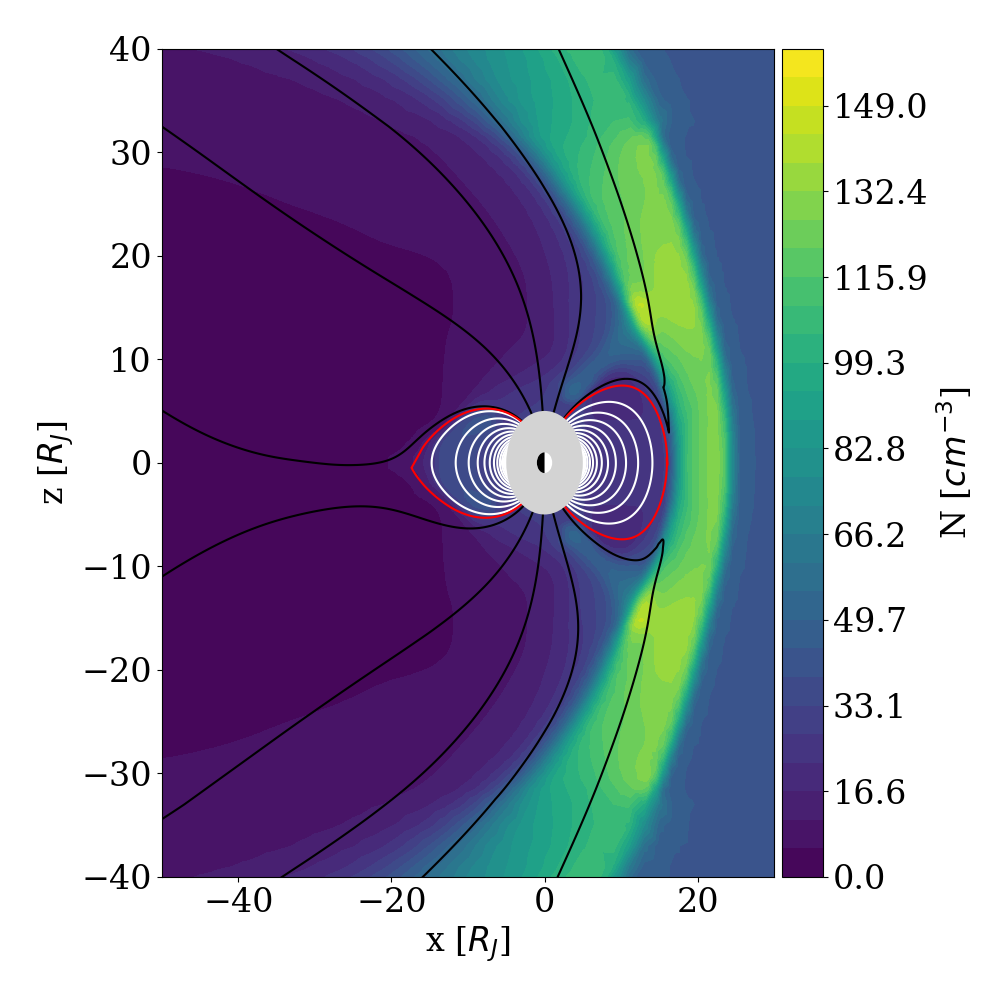}\\
\includegraphics[scale=0.2]{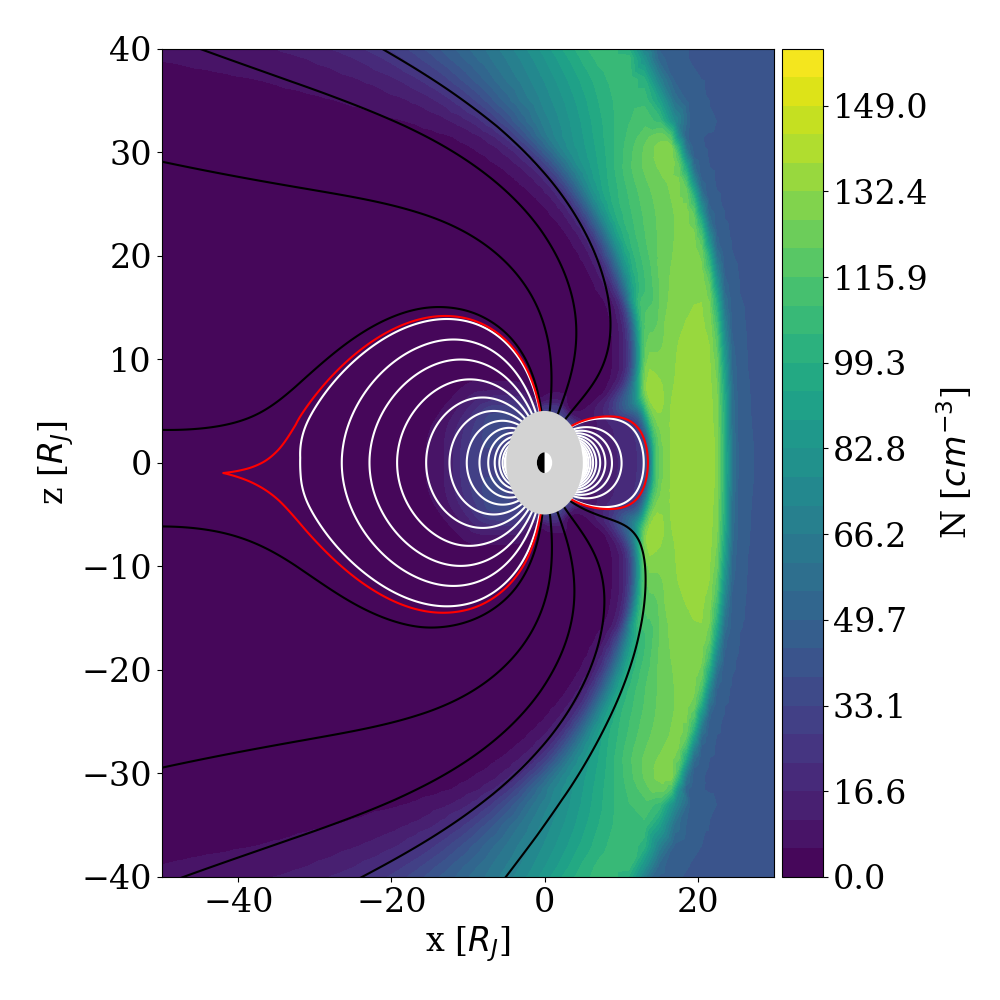}
\includegraphics[scale=0.2]{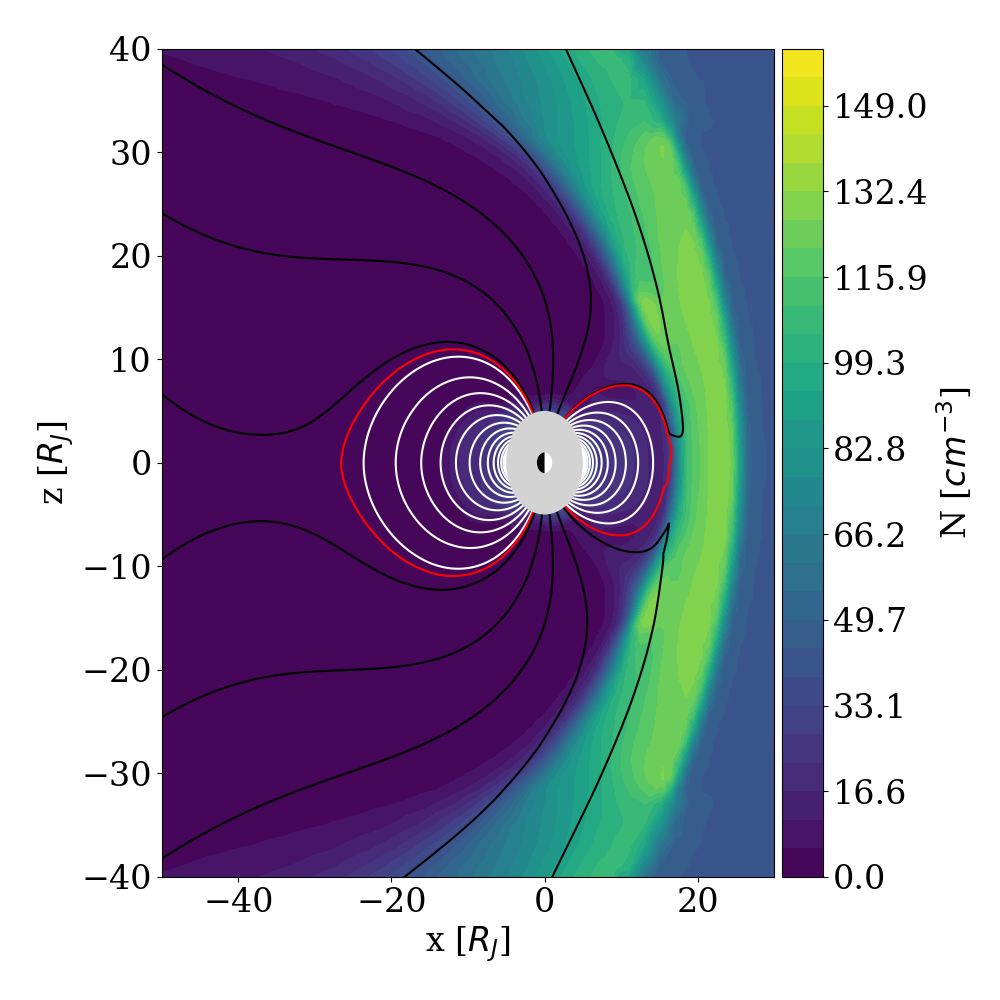}
\caption{Magnetic field lines at the fist time step of the simulation of \textit{top-left}- a tidally locked planet, and \textit{top-right}- a fast-rotating planet (10 hours rotation period); and after 8 hours of running simulation of a \textit{lower-left}- tidally locked planet, and \textit{lower-right}-fast-rotating planet.
\label{fig:tid_vs_rot_magLines_0}}
\end{figure}

\begin{figure}
\centering
\includegraphics[width=80mm]{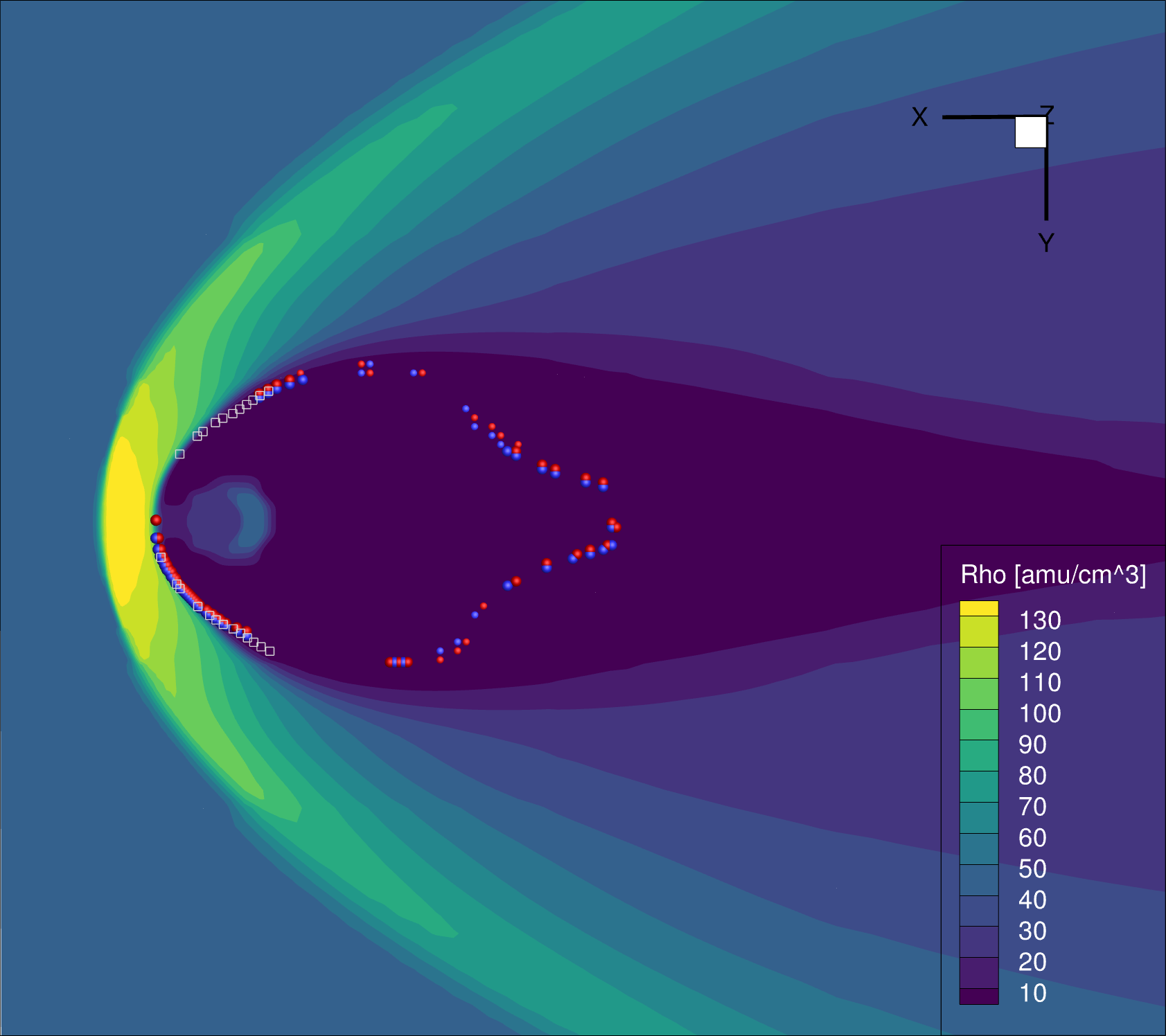}
\includegraphics[width=80mm]{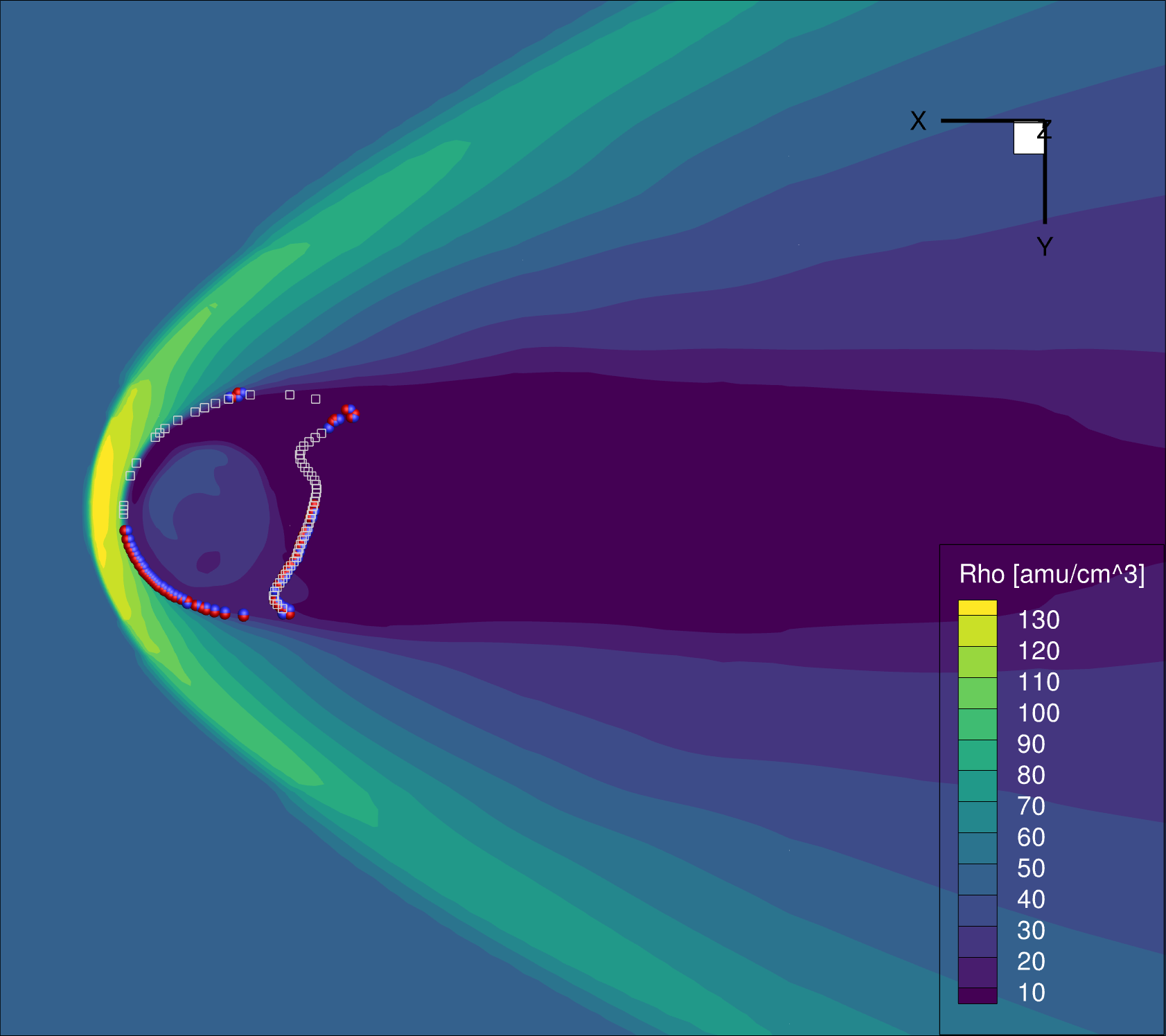}
\caption{The location of magnetic nulls and separators for \textit{right}- a tidally locked planet, and \textit{left}- a fast-rotating planet (10 hours rotation period). The scale and size of Earth's magnetosphere are consistent across both panels in the figure.
\label{fig:tid_vs_rot_3d}}
\end{figure}

\begin{figure}
\centering
\includegraphics[width=80mm]{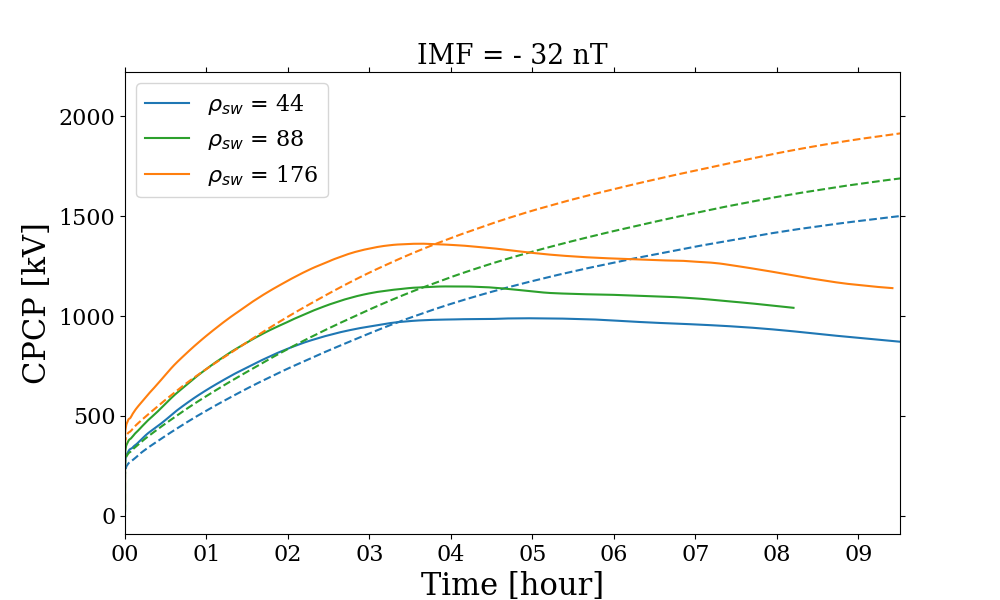}
\includegraphics[width=80mm]{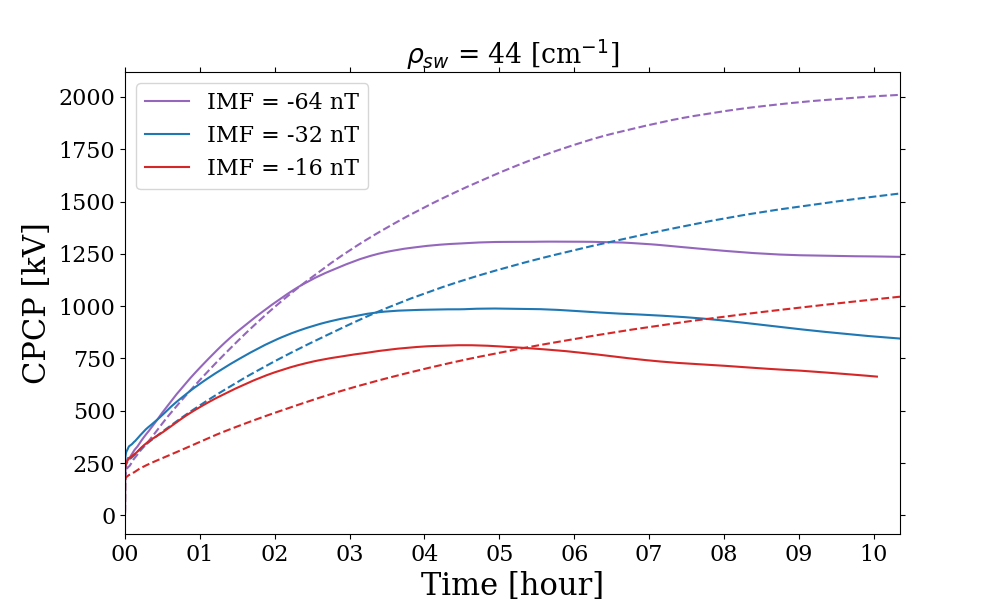}
\caption{CPCP for fast-rotating planets (solid lines) and tidally locked planets (dashed lines) under varying \textit{left}- solar wind density and \textit{right}- solar wind IMF. All other parameters are consistent with those in Table \ref{tab:sw04au}.
\label{fig:swParameters}}
\end{figure}

\noindent A potential explanation for the disparity in CPCP values between tidal and rotating planets lies in variations of the energy dissipation mechanism in these cases. In the case of rapidly rotating gas giants like Jupiter and Saturn, the centrifugal force due to their fast rotation plays a significant role in reducing the dayside reconnection rate \citep{hill1984rotationally}. On rotating planets, the momentum equation for the plasma flow in the magnetosheath, in the planet's rotating frame of reference, is given by:
\begin{equation}
\rho \frac{d \vec{v}}{dt} = \rho \vec{v}.\nabla{v}=
\vec{J} \times \vec{B} -\nabla P + \rho~\Omega^2 ~\vec{r} + 2 \rho ~\vec{v} \times \vec{\Omega}~, \label{eq:momentum}
\end{equation}
where $\rho$ is the plasma flow in the magnetosheath. The centrifugal force caused by the planet's rotation pushes the plasma sunward, opposing the pressure force from the solar wind. This causes the magnetopause to shift further away from the planet. Since the position of the magnetopause is anti-correlated with the dayside reconnection rate \citep{kim2024relation, borovsky2008determines}, a more distant magnetopause leads to lower reconnection rates on the dayside. As a result, the increased centrifugal force on rapidly rotating planets reduces the dayside reconnection rate, which in turn leads to a smaller CPCP.
The analogy here is similar to the force-balance model used to describe the energy input in the Earth’s magnetosphere during low Mach number storms \citep{lopez2010role, lopez2016integrated, bagheri2022solar}. The additional terms related to the planet’s corotation in the right-hand side of Eq. \eqref{eq:momentum} increase the divergence of the magnetospheric flow ($d\vec{v}/{dt} = \rho \vec{v}.\nabla{v}$) on the left-hand side and redirect the flow toward the dawn and dusk regions.  As a result, the geoeffective length decreases, which also leads to a reduction in the dayside reconnection rate. This is illustrated in Figures \ref{fig:tid_vs_rot_3d} and \ref{fig:tid_vs_rot_magLines_0}. Figure \ref{fig:tid_vs_rot_magLines_0} shows the magnetic field lines at the first time step of simulations for a tidally locked planet (top-left panel) and a fast-rotating planet (top-right panel). The bottom panels show the distribution of magnetic field lines after 8 hours of simulation. As seen, the magnetopause is pushed farther away from the fast-rotating planet compared to the tidally locked planet, resulting in a larger dayside region for fast-rotating planets. 
\\
\\
\noindent Furthermore, on fast-rotating planets (assuming rotation from dawn to dusk), the magnetospheric dynamics differ significantly than tidally locked planets. On the dawn side, the magnetospheric plasma moves sunward, opposing the solar wind flow in the magnetosheath. This interaction drags the magnetospheric plasma, creating a return flow deeper within the magnetosphere, which forms a cell-like circulation pattern. These viscous-like interactions extend through the magnetic field to the ionosphere, influencing plasma flow patterns there. On the dusk side, however, the magnetospheric plasma moves anti-sunward, in the same direction as the magnetosheath plasma. This results in lower shear velocities and less intense viscous-like interaction compared to the dawn side. The magnetic shear generated on the dawn side leads to increased plasma density in that region. The shear-induced flows help spread plasma across different parts of the magnetosphere, which increases the divergence of the plasma flow, especially near the magnetopause. According to the Cassak-Shay formula, the rate of dayside reconnection at the magnetopause is closely related to the plasma mass density near the reconnection site \citep{borovsky2008determines}. Therefore, the increased magnetic shear, which raises the divergence of the flow in the magnetosphere, results in a reduced reconnection rate on fast-rotating planets.
\\
\\
Centrifugal force also plays a significant role in nightside reconnection. In the case of Earth, the exchange of mass and energy at the boundary between the solar wind and the magnetosphere, as well as within the magnetodisc on the nightside, forms what is known as the Dungey cycle \citep{dungey1961interplanetary}. However, in fast-rotating planets, the centrifugal force acts on the magnetic field lines in the magnetosphere, causing them to stretch and become distorted as the planet rotates. When the magnetic field lines reach a critical point of tension, they can undergo reconnection, releasing stored magnetic energy and accelerating charged particles as seen in fast-rotating planets in our solar system \citep{guo2018rotationally, guo2019long, kivelson2005dynamical, vogt2020magnetotail}. Magnetic reconnection driven by the centrifugal force linked to planetary rotation is termed Vasyliunas reconnection \citep{vasyliunas1983plasma}.
Figure \ref{fig:tid_vs_rot_3d} illustrates the locations of magnetic reconnection and separation for a tidally locked planet (right panel) and a rotating planet (left panel). The magnetic null points were located using the code from \citep{glocer2016separator}. In the tidal case, magnetic reconnection occurs symmetrically, whereas in the rotating case, symmetry is disrupted due to the planet's rotation. The lack of Vasyliunas reconnection in a tidally locked planet can result in the elongation of its magnetospheric tail and the accumulation of magnetic field lines within it. 
\\
\\
\\
\\
\noindent In summary, on a tidally locked planet, magnetic reconnection events primarily take place on the dayside and at closer distances. In contrast, for a rotating planet, the dayside reconnection rate is lower and occurs at farther distances, while, nightside reconnection, driven by both Dungey and Vasyliunas processes, happens at closer distance, allowing for a greater influx of plasma into the magnetosphere-ionosphere system. This results in a more balanced distribution of magnetic energy between the dayside and nightside compared to tidally locked planets, though there is asymmetry toward dawn and dusk due to the corotation of the planet.
\\
\\
The key question is whether changes in solar wind parameters can diminish the difference in CPCP values between tidally locked and fast-rotating planets. To address this, we performed several simulations. First, we varied the solar wind density while keeping other solar wind parameters constant. Then, we changed the IMF, while other parameters are as those in Table \ref{tab:sw04au}. As shown in Figure \ref{fig:swParameters}, the differences between tidally locked and fast-rotating planets remain significant across all simulations conducted over a 10-hour period. In the left panel of Figure \ref{fig:swParameters}, we observe that increasing solar wind density raises the CPCP values, as expected, but also amplifies the disparity between the two cases of tidally locked (dashed lines) and fast-rotating planets (solid lines). Similarly, the right panel illustrates that increasing the IMF further highlights the differences between the tidally locked and fast-rotating cases.

\section{Exploring the Interplay of Corotation in CPCP}
\begin{figure}[ht!]
\centering
\includegraphics[width=80mm]{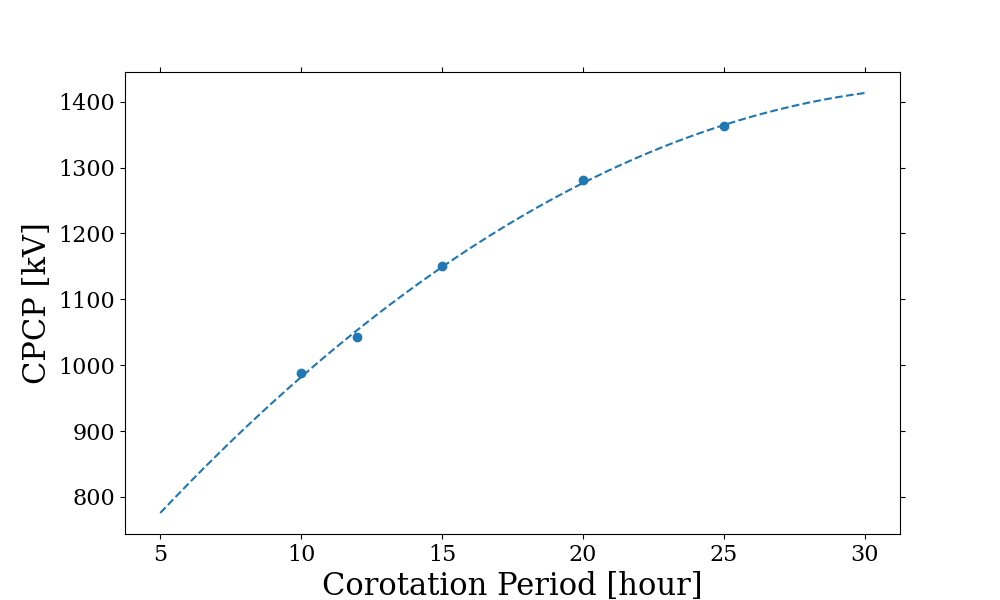}
\includegraphics[width=80mm]{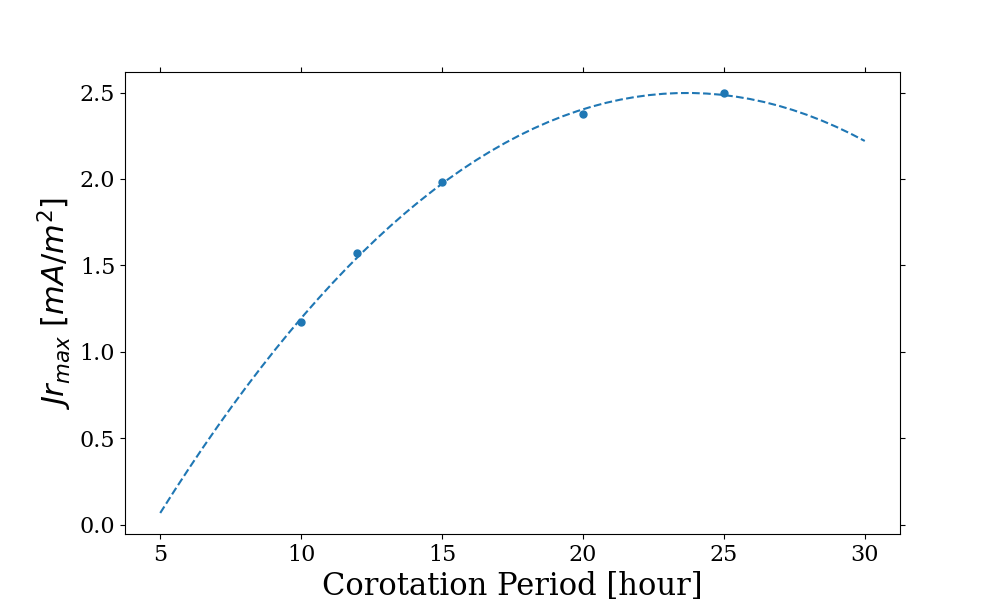}
\caption{\textit{left}-CPCP and \textit{right}-maximum of radial component of the ionospheric current vs. corotation period for a Jupiter-like planet. The solar wind parameters used in these simulations are summarized in Table \ref{tab:sw04au}. The dashed lines represent the best fit to the data.
\label{fig:CPCP_vs_omega}}
\end{figure}

\noindent In this section, we investigate how the CPCP varies with the corotation of the planet. We model a Jupiter-like exoplanet, similar to our previous section, examining corotation periods of 10, 12, 15, 20, and 25 hours. 
As illustrated in the left panel of Figure \ref{fig:CPCP_vs_omega}, we find that as the corotation period decreases, the CPCP also decreases.  
This indicates a systematic relationship between corotation and CPCP, highlighting the influence of the planet’s rotation on its magnetospheric dynamics. Furthermore, as seen in the right panel of Figure \ref{fig:CPCP_vs_omega}, faster corotation is shown to reduce the energy input to the ionosphere, including the convection electric field which is associated with heating the upper atmosphere through Joule heating, and a reduction in the field aligned current that drives the electron beam which makes the aurora \citep{knight1973parallel}. As a result, we expect that corotation may also reduce the strength of the precipitating auroral electron beam, which in turn provides the free energy for the radio emission from the Electron-Cyclotron Maser Instabilities (ECMI). Therefore, we surmise that the ECMI may also be sensitive to planetary corotation which implies that tidally locked planets emit more radiation compared to rotating planets with the same magnetic dipole strength. 

\section{EXPLORING The Role of Planetary Radius}

\begin{figure}
\centering
\includegraphics[width=100mm]{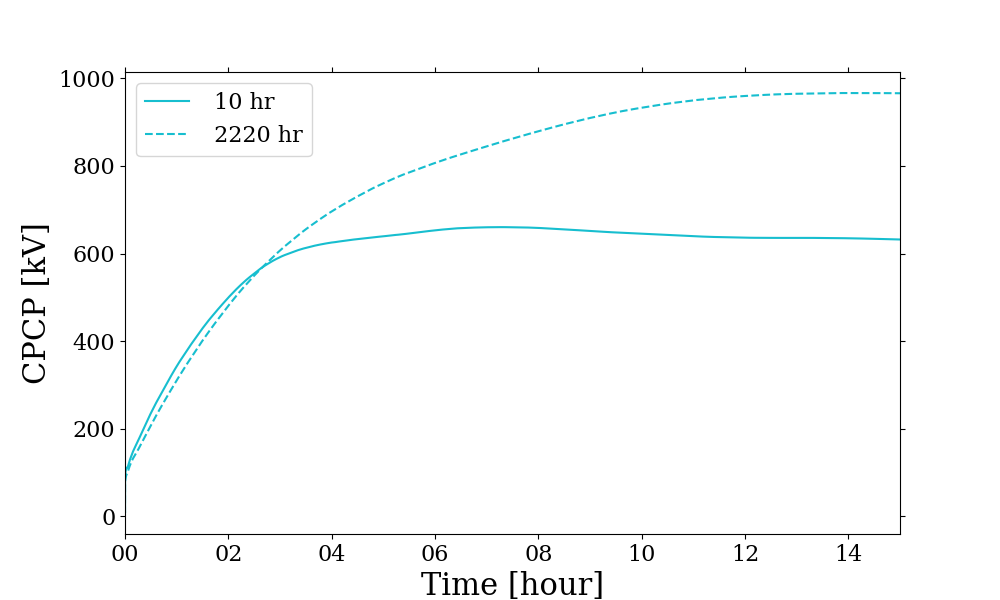}

\caption{CPCP for a tidally locked Saturn-like planet at 0.4 au (2220 hours corotation period) and a fast-rotating planet (10 hours) in 15 hours of simulations. The solar wind parameters used in these simulations are summarized in Table \ref{tab:sw04au}. 
\label{fig:CPCP_saturn}}
\end{figure}

\noindent The difference in CPCP between tidally locked and rotating planets can be attributed to the centrifugal force. This suggests that the planetary radius influences the effects of corotation, as centrifugal force depends on the planet's size. To explore how planetary radius affects the trend seen in CPCP, we repeat the analysis for an Earth-like planet located at 0.4 AU, using solar wind parameters listed in Table \ref{tab:sw04au}. The results show no significant difference in CPCP when varying the corotation speed, even for a 10-hour corotation period. For conciseness, the corresponding results are not presented here. This outcome is consistent with the GAMERA model results \citep{bagheri2024mhd}. Next, we conduct a similar analysis for a Saturn-like planet, which has a radius 9.1 times that of Earth, while its magnetic dipole at the planet's surface is much weaker than Earth's. Figure \ref{fig:CPCP_saturn} compares the CPCP values between a tidally locked Saturn-like planet and a fast-rotating Saturn-like planet (with a 10-hour corotation period). Similar to the Jupiter-like cases, we observe that CPCP values continue to increase in the tidally locked case, whereas the fast-rotating planet reaches equilibrium much more quickly. Since Saturn's magnetic dipole is significantly weaker than Jupiter's, but its radius is comparable, we can conclude that the size of the planetary system influence the effects of corotation which supports our explanation that the centrifugal force is responsible for the observed trend in CPCP values; in other words, reducing the planetary radius and magnetic dipole, like in the Earth-like case, diminishes the observed trend in CPCP.\\ 
 
\section{Exploring the tidal lock effect on the hot-Jupiter systems}\label{sec:hot_jupiter}

\noindent As discussed in \citep{bagheri2024mhd}, MHD simulations often encounter significant gaps and limitations when modeling the magnetospheres of giant exoplanets, like Jupiter, that orbit close to their stars. Their proximity to their host stars can lead to unique magnetospheric structures, complicating the simulations by either requiring the star to be included in the simulation grid or potentially violating MHD boundary conditions. To achieve more accurate results, it would be ideal to simulate both the planet and the star together. However, since this work is one of the first to investigate the effects of tidal locking in hot-Jupiter systems, we assume that our Jupiter-like planet is subjected to an extreme solar wind. Therefore, instead of modifying the MHD framework itself, we focus on adjusting the solar wind inputs and the corotation period for this tidally locked scenario. The solar wind parameters utilized in this section are from \citep{johnstone2015stellar} and summarized in Table \ref{tab:hot_jupiter}.\\
\\
The solar wind parameters used in this section are based on an orbital distance of 0.05 au, which results in a corotation period of approximately 4 days (or 96 hours) for tidally locked planets. At this distance, the planetary conductance may vary. To account for changes in planetary conductance, we perform two sets of simulations for both super-Alfvénic and sub-Alfvénic conditions. The first set corresponds to a fast-rotating and a tidally locked planet with a Pedersen conductance of 105, consistent with previous sections. In the second set, we adjust the Pedersen conductance using Equation 27 from \citep{nichols2016stellar}, resulting in a conductance of $\Sigma_p = 7926$  mho for Jupiter-like planets at an orbital distance of 0.05 au.
\begin{center}
{Key Input Parameters for Simulations in Section \ref{sec:hot_jupiter}}
\begin{tabular}{lllcc}
\hline\hline
corresponds to the orbital distance [au]	&  0.05 \\
& super-Alfvénic & sub-Alfvénic \\
\hline
$\rho_{sw} ~[\text{cm}^{-3}]$	&	3000    &  1000 \\
$v_{sw} ~[\text{km/s}]$	    	&	250	    &  250  \\
$B_{\text{IMF}} ~[\text{nT}]$	&	300	    &  300  \\
tidal corotation period [hour]  &	96		&  96   \\
\hline
\end{tabular} 
\label{tab:hot_jupiter}
\\
\end{center}

\begin{figure}[ht!]
\centering
\includegraphics[width=80mm]{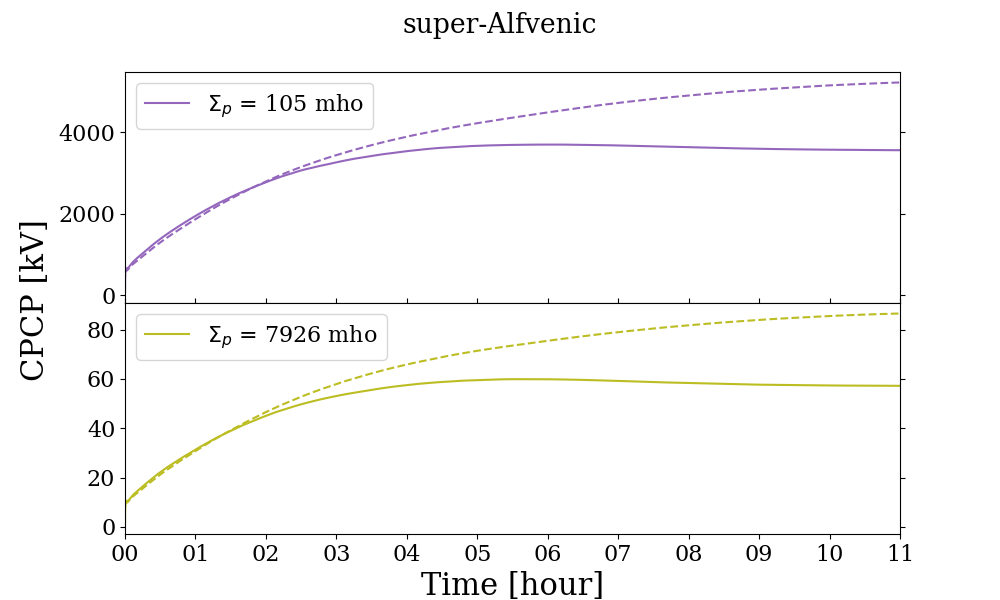}
\includegraphics[width=80mm]{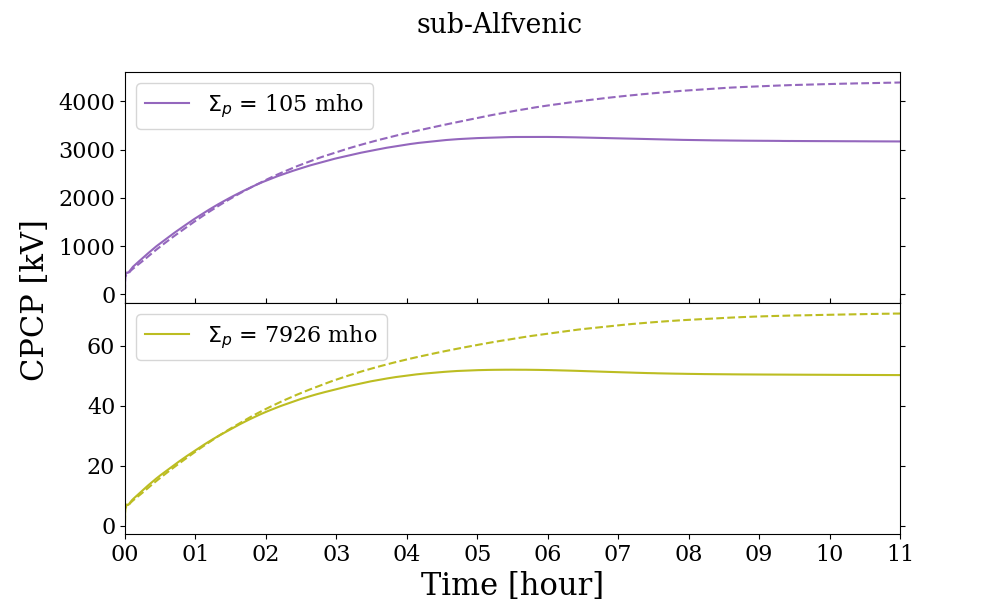}
\caption{CPCP for a tidally locked planet at 0.05 au (4-day corotation period, dashed lines) compared to a fast-rotating planet (10-hour period, solid lines). Results are shown over 11 hours of simulations with varying Pedersen conductance, in the \textit{left-} super-Alfvénic and \textit{right-} sub-Alfvénic zones.
\label{fig:tid_vs_rot_cpcp_0.05}}
\end{figure}

\noindent We investigate the tidal locking effect under two conditions: super-Alfvénic and sub-Alfvénic solar wind. As illustrated in Figure \ref{fig:tid_vs_rot_cpcp_0.05}, the CPCP exhibits the same trends in both scenarios; while the CPCP values increase significantly, the overall trend comparing rotating and tidally locked planets remains consistent. This indicates that the influence of corotation is still significant even under the extreme conditions of a high interplanetary magnetic field and solar wind density. Also, the trend is independent of the choice of Pederesen conductance in the ionosphere.\\
\\
As planets orbit closer to their host star, their corotation period decreases, making the CPCP behavior more similar to that of fast-rotating planets. For example, the corotation period of planet NGTS-10 b is approximately 18.5 hours \citep{2020MNRAS.493..126M}. In this case, even though the hot-Jupiter may be tidally locked, its CPCP trends align more closely with those of fast-rotating planets.

\section{Conclusion}

\noindent In this study, we explore how corotation affects the magnetosphere-ionosphere system of exoplanets and its interaction with stellar winds. Using the SWMF MHD code, we simulate the magnetosphere of a Jupiter-like planet under various conditions, including different solar wind parameters and corotation speeds. Our findings reveal that the maximum CPCP of a tidally locked planet is approximately twice that of a fast-rotating planet. This result corroborates with our previous investigation utilizing the GAMERA MHD code \citep{bagheri2024mhd}. We propose that the difference in CPCP is due to the influence of centrifugal force. In fast-rotating planets, the centrifugal force pushes the magnetopause farther from the planet, reducing the dayside reconnection rate and lowering CPCP. In contrast, tidally locked planets lack this centrifugal force, which keeps the magnetopause closer to the planet and leads to higher CPCP. Furthermore, the absence of Vasyliunas reconnection in the magnetotail of tidally locked planets enhances the asymmetry between dayside and nightside reconnection, further distinguishing their behavior from fast-rotating planets.\\
\\
We also explore how changes in corotation speed affect the CPCP. Our findings show that as we increase the corotation speed, the CPCP decreases in a linear relationship. However, for Earth, considering different corotation period from 6 hours to 24 hours, the effect of corotation on the CPCP is minimal. This suggests that the size of a planet is crucial in understanding the influence of corotation within the magnetosphere-ionosphere system. For smaller planets changes in corotation speed have little impact on how the magnetosphere interacts with stellar winds. This finding aligns with the results in \citep{bagheri2024mhd}, where a tidally locked Earth-like planet showed no discernible difference in CPCP compared to a rotating one using GMAERA MHD code.\\
\\
Furthermore, we adjust the solar wind parameters to reflect the extreme conditions of hot-Jupiter systems. Our results indicate that even under these conditions, where both the interplanetary magnetic field and solar wind density are high, the influence of corotation remains significant. \\
\\
This discovery has important implications for our understanding of potential of radio emissions from tidally locked exoplanets. Traditionally, these planets were expected to emit minimal radio waves due to their slow rotations. However, our research indicates that if a tidally locked exoplanet possesses a dynamo-generated magnetic field, similar to Earth's, it can actually produce more radio emissions. This increase is attributed to the greater energy present in the magnetosphere-ionosphere systems, which enhances the generation of radio signals.
\\
\\
\section*{acknowledgment}
\noindent We acknowledge the Texas Advanced Computing Center (TACC) at The University of Texas at Austin for providing computational resources that have contributed to the research results reported within this paper (\url{http://www.tacc.utexas.edu}). We also acknowledge the use of the Derecho supercomputer at the National Center for Atmospheric Research (NCAR) \citep{computational2017cheyenne}. A.G. acknowledges support from the RHaPS project, NASA Proposal \#22-ICAR22-2-0030.  F.Bagheri's research was supported by an appointment to the NASA Postdoctoral Program at the NASA Goddard Space Flight Center, administered by Oak Ridge Associated Universities under contract with NASA. 

\bibliographystyle{aasjournal} 
\bibliography{main}




\end{document}